# Conformal Titanium Nitride in a Porous Silicon Matrix: a Nanomaterial for In-Chip Supercapacitors

*Kestutis Grigoras, Jari Keskinen, Leif Grönberg, Elina Yli-Rantala, Sampo Laakso, Hannu Välimäki, Pertti Kauranen, Jouni Ahopelto, Mika Prunnila\**

*VTT Technical Research Centre of Finland, P.O.Box 1000, FI-02044 VTT, Finland*

*\*email: mika.prunnila@vtt.fi*

**Abstract**

Today's supercapacitor energy storages are typically discrete devices aimed for printed boards and power applications. The development of autonomous sensor networks and wearable electronics and the miniaturisation of mobile devices would benefit substantially from solutions in which the energy storage is integrated with the active device. Nanostructures based on porous silicon (PS) provide a route towards integration due to the very high inherent surface area to volume ratio and compatibility with microelectronics fabrication processes. Unfortunately, pristine PS has limited wettability and poor chemical stability in electrolytes and the high resistance of the PS matrix severely limits the power efficiency. In this work, we demonstrate that excellent wettability and electro-chemical properties in aqueous and organic electrolytes can be obtained by coating the PS matrix with an ultra-thin layer of titanium nitride by atomic layer deposition. Our approach leads to very high specific capacitance (15 $Fcm^{-3}$), energy density (1.3 $mWhcm^{-3}$), power density (up to 214 $Wcm^{-3}$) and excellent stability (more than 13,000 cycles). Furthermore, we show that the PS-TiN nanomaterial can be integrated inside a silicon chip monolithically by combining MEMS and nanofabrication techniques. This leads to realisation of in-chip supercapacitor, i.e., it opens a new way to exploit the otherwise inactive volume of a silicon chip to store energy.

# 1 Introduction

Supercapacitors are high capacitance density electrochemical double layer capacitors (EDLC) providing a high power energy storage [1,2] that could be utilized also in miniaturized devices [3-11]. Integration of a small volume supercapacitor with the active devices will largely boost the miniaturization and advance the power efficiency of, for example, capacitive energy harvesters [12] and solar cells [13]. In discrete supercapacitor devices typically high surface area carbon materials, e.g. activated carbon, carbide derived carbon, carbon nanotubes or graphene, are used as the electrodes [1,2]. Carbon based materials have been also investigated for on-chip integration. In this respect, e.g., photoresist derived carbon [3-7], graphene reduced from graphite oxide [8-10], silicon carbide-derived carbon films [11] and carbon nanotube forest devices [14] have been considered. Best performance numbers for on-chip type devices have been found from graphene oxide/reduced graphene oxide (GO/RGO) microcapacitor with volumetric and areal specific capacitance (8 μm thick electrodes) of 2.35 $Fcm^{-3}$ and 2 $mFcm^{-2}$, respectively, and 200 $Wcm^{-3}$ power density and 2 $mWhcm^{-3}$ energy density [9]. Devices with silicon carbide-derived carbon films have been also demonstrated [11] with areal capacitance density up to 0.7 $mFcm^{-2}$, which is similar to the performance of CNT based electrodes (0.4 $mFcm^{-2}$) [14]. A shortcoming of this technique regarding integration is the relatively high temperature needed for graphitization.

Silicon has received significantly less attention than carbon based materials as a supercapacitor material although, for example, porous silicon (PS) provides a very high surface area matrix with relatively well controlled and reproducible properties. Silicon nanowires (SiNW) can be addressed to the same group. However, porous silicon and Si NWs, at large part, have some challenges as electrode material, such as chemical stability, relatively high resistivity and poor wettability [15-20]. Coating or doping of Si structures has been recognized as an attractive route to reduce the resistance and increase the stability [17-17, 20]. Highly doped SiNWs can exhibit excellent voltage cycling stability, but they offer few orders of magnitude lower capacitance densities in comparison to carbon based devices [18, 19]. Also for coated electrodes the performance has still been significantly below that of carbon on-chip supercapacitors [9], because in the end the power density is detrimentally limited by the high resistance of the complex Si nanostructure. Areal capacitance densities obtained for PS and doped SiNW based devices are few hundreds of $\mu F/cm^2$ [15, 16, 20].

In this work, we demonstrate that by coating PS by conformal ultra-thin titanium nitride (TiN) layer highly conductive and wettable and stable supercapacitor electrodes in aqueous and organic electrolytes can be obtained. Conformal coating of PS is not straightforward due to very high aspect ratios which can reach 1:1000 and above. Homogeneous thin film of TiN can, however, be deposited inside the PS matrix by atomic layer deposition (ALD) process that is optimized for extremely high aspect ratio structures. We show that PS matrix with a ~10 nm thick layer of TiN deposited by optimized ALD enables fabrication of very efficient supercapacitor electrodes with almost ideal EDLC characteristics. This approach basically combines the good electrochemical properties of TiN [21] and large area of the PS matrix. One of the key results here is that the use of the PS-TiN electrodes enables the realization of an efficient supercapacitor integrated inside the silicon wafer – the in-chip supercapacitor. Silicon technology is still typically planar technology and leaves the bulk of the silicon wafer merely as a support with relatively large unused volume. We demonstrate that the volume of a bulk wafer or the handle of a SOI wafer can be effectively used to embody a PS-TiN supercapacitor providing a route to new type of integrated energy storages.

## 2 Experimental

PS supercapacitor electrode characterization was performed using assembled test devices (Figure 1), which provide versatile platform for material testing. Heavily doped 150 mm diameter Si wafers were used as initial material. A 200 nm thick silicon nitride layer was deposited by low pressure chemical vapour deposition (LPCVD) technique to form a mask for porous silicon formation. Silicon nitride mask was patterned by UV lithography and plasma etching, resulting in 20 identical circle-shaped windows of 1.4 cm diameter each. Porous silicon wells were prepared by electrochemical etching of a whole silicon wafer in commercial etching cell (AMMT, Germany) containing 50% HF and ethanol solution 1:4. The porosity of the PS layer, as evaluated from gravimetric measurements (Sartorius CP224S laboratory balances) [22], was 87-88% and the pore diameter was about 100 nm (Figures 1b, c). The mass of single porous electrode layer (6 µm thick, 1.5 cm$^2$ area) was evaluated to be 0.27 mg.

The TiN coating of the pores was performed by ALD, which can provide extremely conformal layer even inside structures with a very high aspect ratio [23]. The thermal ALD process was conducted in a Beneq TFS-500 reactor at 450$^o$C temperature keeping 800 Pa pressure inside the reaction chamber and using TiCl$_4$ and ammonia as precursors [24,25]. Nitrogen was used as a carrier gas for precursor transportation and also for the purging of the reaction chamber after each precursor pulse. The precursor pulse/purge duration was 0.5 s / 10 s and 2 s / 40 s for TiCl$_4$ and ammonia, respectively. Process sequence is depicted schematically in Figure 1d. The formation of a conformal ~10 nm thick layer was confirmed by scanning electron microscopy (SEM) (Figure 1c). Amount of TiN layer deposited inside the pores of one electrode was 0.9 mg, as estimated from gravimetrical measurements.

Processed wafers were cleaved into 23x23 mm$^2$ chips with a single porous area at the centre. Two silicon chips with porous silicon area in the middle were sandwiched with a frame of polydimethylsiloxane (PDMS) between them (Figure 1a). The 2 mm-thick PDMS frame was prepared from Sylgard 184 and it served as a reservoir for the electrolytes. Silicon chips and PDMS were joined together by bonding [26]. The cell was filled with electrolyte using a syringe. We used 1 M NaCl water solution (aqueous electrolyte) and 0.5 M TEABF4 in PC (tetraethyl ammonium tetrafluoroborate in propylene carbonate, organic electrolyte). The complete parameters for devices #A (aqueous electrolyte) and #O (organic electrolyte) can be found in Supplementary Data (Table S1).

The in-chip supercapacitor device consists of two TiN coated porous Si electrodes inside Si chip with electrolyte reservoir in between. The device fabrication is illustrated in Figure 2. The in-chip device fabrication started with ALD aluminium oxide layer (15 nm) and silicon nitride layer (1.5 um thick) deposition which served as an etch stop layer and supporting structure for the electrodes, respectively (Figure 2a). Contact openings were done through both layers (Figure 2b) and supercapacitor electrode contacts were prepared by sputtering and patterning of aluminium (figure 2c). Aluminium oxide was prepared by ALD on the top side of the wafer and patterned to act as a mask for anisotropic plasma etching through the wafer to form trench reservoirs for the electrolyte (Figure 2d). The next steps were porous silicon formation on the trench sidewalls and TiN coating by ALD (Figure 2e). These two steps were almost identical as those used for a single porous electrode preparation described above. Due to the conformal growth, provided by ALD, the TiN layer creates a short circuit at the bottom of the trench. The short circuit was removed by ion beam etching resulting in supercapacitor device where two galvanically separated supercapacitor electrodes with electrolyte reservoir are inside silicon chip – the in-chip supercapacitor. The electrolyte reservoir was filled with aqueous electrolyte (1M NaCl).

Galvanostatic charge-discharge curves were measured with Arbin supercapacitor test station. In the cyclic voltammetry (CV) scans and electrochemical impedance spectroscopy (EIS) we utilized IviumTech potentiostat.

# 3 Results and discussion

In high surface area/aspect ratio silicon nanostructures the resistance increases due to traps and depletion effects [27] and it has been shown that even highly doped PS behaves like a low loss dielectric instead of a conductor [28]. The resistivity of our TiN layer was 0.16 mOhm-cm, which is comparable to data reported in literature for thin TiN films [24, 29]. The TiN coating solves the resistance issue of the PS matrix and also affects favourably on the wettability of the electrodes (Figure 3). Pristine PS is clearly pronounced hydrophobic surface (surface angle about 142$^o$) (Figure 3a). Whereas, TiN coating leads to hydrophilic surface with surface angle of about 20$^o$ (Figure 3b). With organic electrolyte on pristine PS we obtained a surface angle of approximately 22$^o$ (Figure 3c) and by adopting the TiN coating the surface angle drops to below 10$^o$ (Figure 3d). In fact, after the TiN coating the contact angle is difficult to evaluate accurately because in less than a second the droplet is effectively adsorbed into the pores (Figure 3e).

Galvanostatic charge-discharge curves measured at 1.0 mA (0.67 mAcm$^{-2}$) for devices with both types of electrolytes are shown in Figure 4a. The triangular, almost symmetric shape of the curves is a signature of good performance as EDLC. The efficiency of the supercapacitors as evaluated from the charge-discharge curves is high, 88% and 83% for the aqueous and the organic electrolyte, respectively. Self-discharge tests show that the electrodes lose 67 % (50 %) of their charge in ~19 (~4.5) hours (Supplementary Data Figure S1). The galvanostatic charge-discharge characterization of devices with as prepared porous silicon electrodes without the TiN coating show poor performance (Supplementary Data Fig. S2a), consistent with previous literature on PS and SiNW devices [15-18, 20].

Small signal frequency response (Nyquist plot) of the PS-TiN supercapacitors is shown in Figure 3b. Almost vertical angle of the curve indicates almost purely capacitive behaviour of the devices. The semicircle in the high frequency part (Z' and Z'' approaching to minimum, Figure 4b Inset) is very small, confirming small resistance of diffusion of ions in electrolyte inside the pore to- and from the electrode surface [30, 31]. Equivalent series resistance (ESR) values evaluated from the Nyquist plot (by extrapolating the straight part of the curve to the intersect with the x-axis [32]) are 5 Ω and 17-18 Ω for samples #A and #O, respectively. The ESR is mainly caused by the electrolyte resistance arising from the 2 mm spacing between the PS chips (Figure 1a). Intrinsic ESR can be estimated considering a smaller distance between the chips. Reducing the distance to 100 μm will result in decrease of ESR to 3.6 Ω and 4.0 Ω for samples #A and #O, respectively.

Extremely good performance of the PS-TiN electrodes was further confirmed by extensive cyclic voltammetry (CV) measurements (Figures 4c, d). CV testing up to 13,000 cycles shows very stable behaviour and good capacitance retention for both aqueous (Figure 4c) and organic electrolytes (Figure 4d). The CV curves have almost ideal rectangular shape (Insets of Figures 4c and 4d), whereas the uncoated samples show again highly non-ideal behaviour (Supplementary Data Figures S2b, S3). Measurements with a reference electrode and one supercapacitor electrode give similar characteristics (Supplementary Data Figure S4).

Volumetric power and energy of the PS-TiN electrodes and other approaches can be found from the Ragone plot presented in Figure 5a. The energy and power densities were evaluated using capacitance and ESR values obtained from charge-discharge experiments performed at different discharge current densities (0.67 mA/cm$^2$ and 0.067 mA/cm$^2$). The highest power point in the Ragone plot is the match impedance point calculated for power available for the load at maximum power with ½ of maximum energy when the load equals with ESR (see Supplementary Data for further details). Match impedance point is shown also for the case of 100 µm PDMS and organic electrolyte. The obtained volumetric densities of energy (0.5 mWhcm$^{-3}$ for aqueous and 1.3 mWhcm$^{-3}$ for organic electrolyte) and power (56 Wcm$^{-3}$ for aqueous and 214 Wcm$^{-3}$ for organic electrolyte) compare favourably to those of other approaches. These both quantities are at least two orders of magnitude larger than the corresponding values for doped and coated silicon nanowires [19, 20], TiN nanowires [21] or silicon carbide nanowires [33]. The maximum power density with organic electrolyte is about two orders of magnitude larger than the power density obtained with supercapacitors based on porous silicon reported in literature [15-17]. Areal power and energy density of PS-TiN electrodes also compare favourably to those of other approaches (Figure 5b).

The assembled test structure of Figure 1a with two symmetrical PS electrode chips serves as a versatile material test platform and the ultimate goal here is to demonstrate an integrated supercapacitor in which both electrodes are located inside a single silicon chip. Such in-chip supercapacitor leaves the surface of the Si chip free for active devices. The in-chip supercapacitor was realised by implementing the PS-TiN electrodes onto the sidewalls of deep vertical trenches etched through a silicon wafer and situating aluminium contacts on the back side of the chip as described in the Experimental Section (Figure 2). This approach allows us to utilize the bulk of the chip efficiently (large capacitance per foot print) and monolithic in-chip integration of supercapacitor electrodes is achieved for the first time. Micrographs and schematics of fabricated in-chip device are shown in Figures 6a-e.

Performance of the in-chip supercapacitor device was tested by filling the trenches with NaCl electrolyte and performing CV measurements. The results of the CV characterization presented in Figure 5f (inset) confirm the functionality of the in-chip supercapacitor. The shape of CV curve exhibits nearly ideal EDLC character similar to the one measured from the material test assembly (Figure 4c Inset). The retention performance is also good (Figure 5f). Furthermore, the current density (normalized to the vertical surface area of porous electrodes) compares well to those values obtained for the assembled device. In-chip device with only TiN coating but no porous Si was also tested. This device has orders of magnitude lower capacitance in comparison to the PS-TiN device (Supplementary Data Figure S5).

The specific capacitance density of the PS-TiN electrodes with aqueous electrolyte determined from the assembled test devices reaches 10 Fcm$^{-3}$ (Supplementary Data Table S1). Calculated specific capacitance density of the in-chip device (~15 F/cm$^3$) compares well to this value. However, the most important figure of merit for the in-chip device is the capacitance per foot print on chip. The PS-TiN in-chip device offers 5 mF/cm$^2$ capacitance per foot print exceeding the values obtained for on-chip devices based on photoresist derived carbon (1 mF/cm$^2$ [7]), graphitization of polycrystalline silicon carbide (0.7 mF/cm$^2$ [11]), or laser reduced graphene oxide (2 mF/cm$^2$ [9]). It should be noted that the in-chip electrode design here is relative loose: a foot print of the device (Figure 5d) is 0.25 cm$^2$, whereas the area of one electrode (the sidewall with porous silicon, Figure 5b) is 0.9 cm$^2$. Higher capacitance per foot print can be reached by optimizing the shape, width and separation of the PS-TiN electrodes, i.e., by increasing the density of comb like interdigitated structures beyond the proof-of-concept device reported here. Increase of the thickness of the PS layer will also directly affect capacitance values. For example, 20 µm thick PS (with the same electrode configuration) instead of 5 mF/cm$^2$ would result in 50 mF/cm$^2$ capacitance per foot print. Such a high capacitance density value suggests that PS-TiN in-chip supercapacitors can provide an attractive energy storage solution for various devices and systems including autonomous sensor networks and wearable devices.

# 4 Conclusions

We have demonstrated efficient and stable supercapacitor electrodes by combining two different elements nanotechnology: porous silicon and TiN coating by atomic layer deposition. The coating passivates the surface of chemically instable porous silicon and significantly reduces the overall resistance, leading to power and energy densities comparable with the levels of carbon based materials. We also demonstrated that the volume of a silicon chip can be effectively used to embody single porous Si-TiN supercapacitor with two galvanically decoupled electrodes and the electrolyte. Such in-chip supercapacitor provides a route to new type of miniaturized components and assemblies that require local integrated energy storage.

**Acknowledgements**

Markku Tilli is acknowledged for providing the silicon wafers. Merja Markkanen and Gao Feng are thanked for technical assistance in the sample fabrication. This work has been partially funded by the Academy of Finland through the Atomic Layer Deposition Centre of Excellence project.

**Appendices**

Supplementary Data accompanies this paper.

**Competing financial interests:** the authors declare no competing financial interests.

**References**

1. P. Simon, Y. Gogotsi, Nature Materials 7 (2008) 845-854.
2. C. Zhao, W. Zheng, Front. Energy Res. 3 (2015) doi:10.3389/fenrg.2015.0023.
3. B. Y. Park, L. Taherabadi, Ch. Wang, J. Zoval, M. Madou, J. Electrochem. Soc. 152 (2005) J136-J143.
4. V. Penmatsa, J-H. Yang, Y. Yu, Ch. Wang, Carbon 48 (2010) 4109-4115.
5. M. Beidaghi, W. Chen, Ch. Wang, J. Power Sources 196 (2011) 2403-2409.
6. S. Sharma, M. Madou, Biomimetic and Nanobiomaterials 1 (2012) 252-265.
7. B. Hsia, M. S. Kim, M. Vincent, C. Carraro, R. Maboudian, Carbon 57 (2013) 395-400.
8. M. M. Hantel, T. Kaspar, R. Nesper, A. Wokaun, R. Kötz Chem. Eur. J. 18 (2012) 9125-9136.
9. M. F. El-Kady, R. B. Kaner, Nat. Commun. 4 (2013) 1475 doi:10.1038/ncomms2446.
10. W. Gao, N. Singh, L. Song, Z. Liu, A. L. M. Reddy, L. Ci, R. Vajtai, Q. Zang, B. Wei, P. M. Ajayan, Nature Nanotechnology 6 (2011) 496-500.
11. F. Liu, A. Gutes, I. Laboriante, C. Carraro, and R. Maboudian, Appl. Phys. Lett. 99 (2011) 112104-112106.
12. A. Varpula, S. Laakso, T. Havia, J. Kyynäräinen, M. Prunnila, Sci. Rep. 4 (2014) 6799 doi:10.1038/srep06799.
13. L. Thekkekara, B. Jia, Y. Zhang, L. Qiu, D. Li, M. Gu, Appl. Phys. Lett. 107 (2015) 031105.
14. Y. Jiang, Q. Zhou, L. Lin, Planar MEMS supercapacitor using carbon nanotube forests. Digest Technical papers IEEE MEMS 09 Conf. Sorrento Italy, June 2009, p.587.
15. S. E. Rowlands, R. J. Latham, W. S. Schlindwein, Ionics 5 (1999) 144-149.
16. S. Desplobain, G. Gautier, J. Semal, L. Ventura, M. Roy, Phys. Stat. Sol. C 4 (2007) 2180-2184.
17. L. Oakes, A. Westover, J. W. Mares, S. Chatterjee, W. R. Erwin, R. Bardhan, S. M. Weiss, C. L. Pint, Sci. Rep. 3 (2013) 3020 doi:10.1038/srep03020.

18. F. Thissandier, A. Le Comte, O. Crosnier, P. Gentile, G. Bidan, E. Hadji, T. Brousse, S. Sadki, Electrochemictry Communications 25 (2012) 109-111.
19. F. Thissandier, L. Dupre, P. Gentile, T. Brousse, G. Bidan, D. Buttard, S. Sadki, Electrochimica Acta 117 (2014) 159-163.
20. J. P. Alper, M. Vincent, C. Carraro, R. Maboudian, Appl. Phys. Lett. 100 (2012) 163901-3.
21. X. Lu, G. Wang, T. Zhai, M. Yu, S. Xie, Y. Ling, Ch. Liang, Y. Tong, Y. Li, Nano Lett. 12 (2012) 5376-5381.
22. M. du Plessis, ECS Transactions 9 (2007) 133-142.
23. K. Grigoras, V-M. Airaksinen, S. Franssila, J. Nanoscience Nanotechnol. 9 (2009) 3763-3770.
24. K. Elers, J. Winkler, K. Weeks, S. Marcus, J. Electrochem. Soc. 152 (2005) G589-G593.
25. H. Van Bui, A. W. Groenland, A. A. I. Aarnink, R. A. M. Wolters, J. Schmitz, A. Y. Kovalgin, J. Electrochem. Soc. 158 (2011) H214-H220.
26. K. Haubert, T. Drier, D. Beebe, Lab on Chip 6 (2006) 1548-1549.
27. V. Lehmann, F. Hofmann, F. Moller, U. Grunning, Thin Solid Films 255 (1995) 20-22.
28. P. Sarafis, E. Hourdakis, G. Nassiopoulou, IEEE Trans. El. Dev. 60 (2013) 1436-1443.
29. M. Ritala, M. Leskelä, E. Rauhala, J. Jakinen, J. Electrochem. Soc. 145 (1998) 2914-2920.
30. B. E. Conway, Electrochemical Supercapacitors: Scientific Fundamentals and Technological applications, Kluwer Academic Press/Plenum Publishers, New York, 1999, pp. 380-382, 506-509.
31. M. D. Stoller, S. Park, Y. Zhu, J. An, R. S. Ruoff, Nano Lett. 8 (2008) 3498-3502.
32. M. D. Stoller, R. S. Ruoff, Energy Environ. Sci. 3 (2010) 1294-1301.
33. J. Alper, M. S. Kim, M. Vincent, B. Hsia, V. Radmilovic, C. Carraro, R. Maboudian, J. Power Sources 230 (2013) 298-302.

# Figures and captions

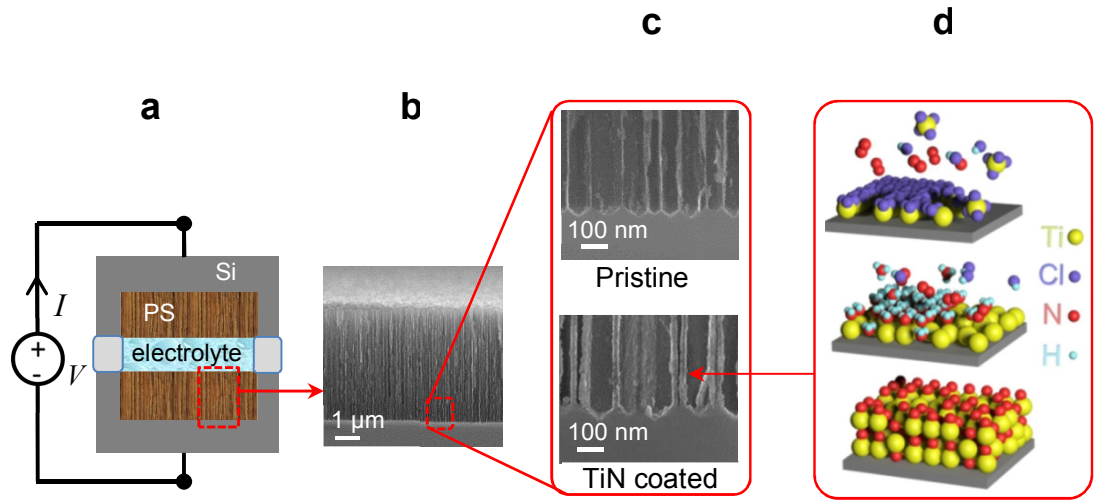

**Figure 1.** Assembled PS-TiN test supercapacitor with separate electrode chips. (a) Schematic cross-section of the assembled supercapacitor device and external circuitry. PDMS collar (light gray) forms the electrolyte reservoir. (b) SEM image of cross-section of porous silicon layer. (c) Larger magnification SEM pictures of bottom part of as prepared PS layer (upper) and TiN conformally coated PS layer (lower). (d) 3D illustration of two ALD cycles of TiN growth.

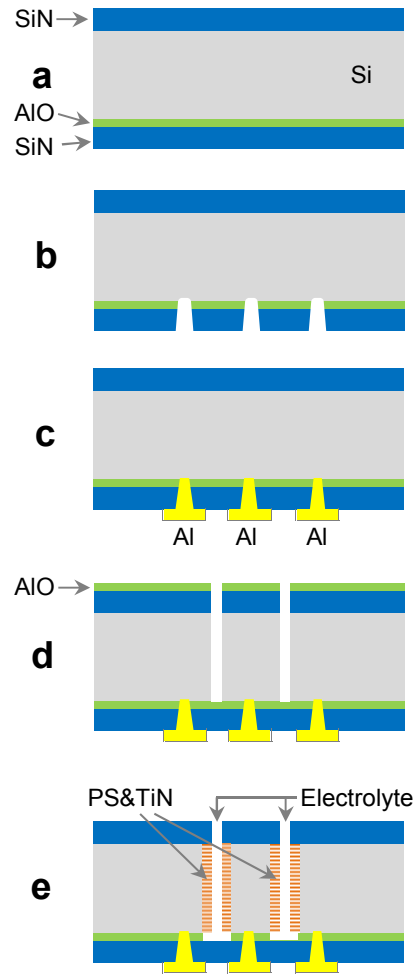

**Figure 2.** Fabrication steps of the in-chip supercapacitor. (a) ALD of $Al_2O_3$ (as through-silicon-etch stop layer) and LPCVD of low stress SiN. (b) Lithography and contact hole. (c) Contact metalization and patterning. (d) Trench side patterning and through wafer plasma etching. (e) Porous silicon formation and TiN deposition inside the pores by ALD. The trenches form the electrolyte reservoir and spacer.

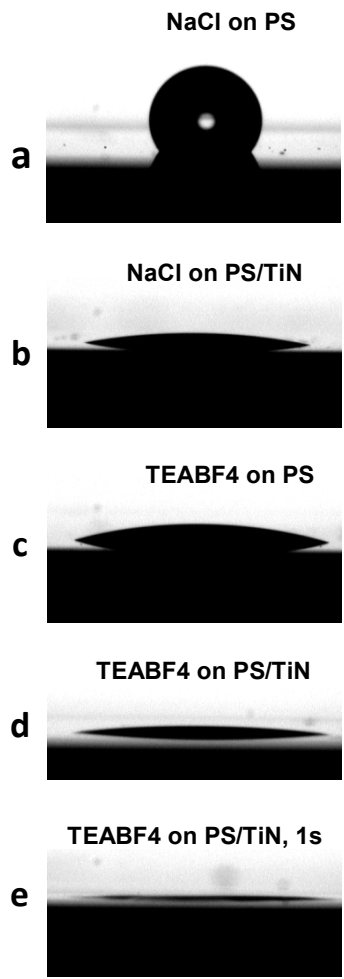

**Figure 3.** Improvement of wettability by TiN coating. Contact angle meter (CAM100) pictures of droplets of (a,b) aqueous electrolyte and (c,d,e) organic electrolyte. (a, c) Pristine porous silicon and (b, d, e) TiN coated porous silicon.

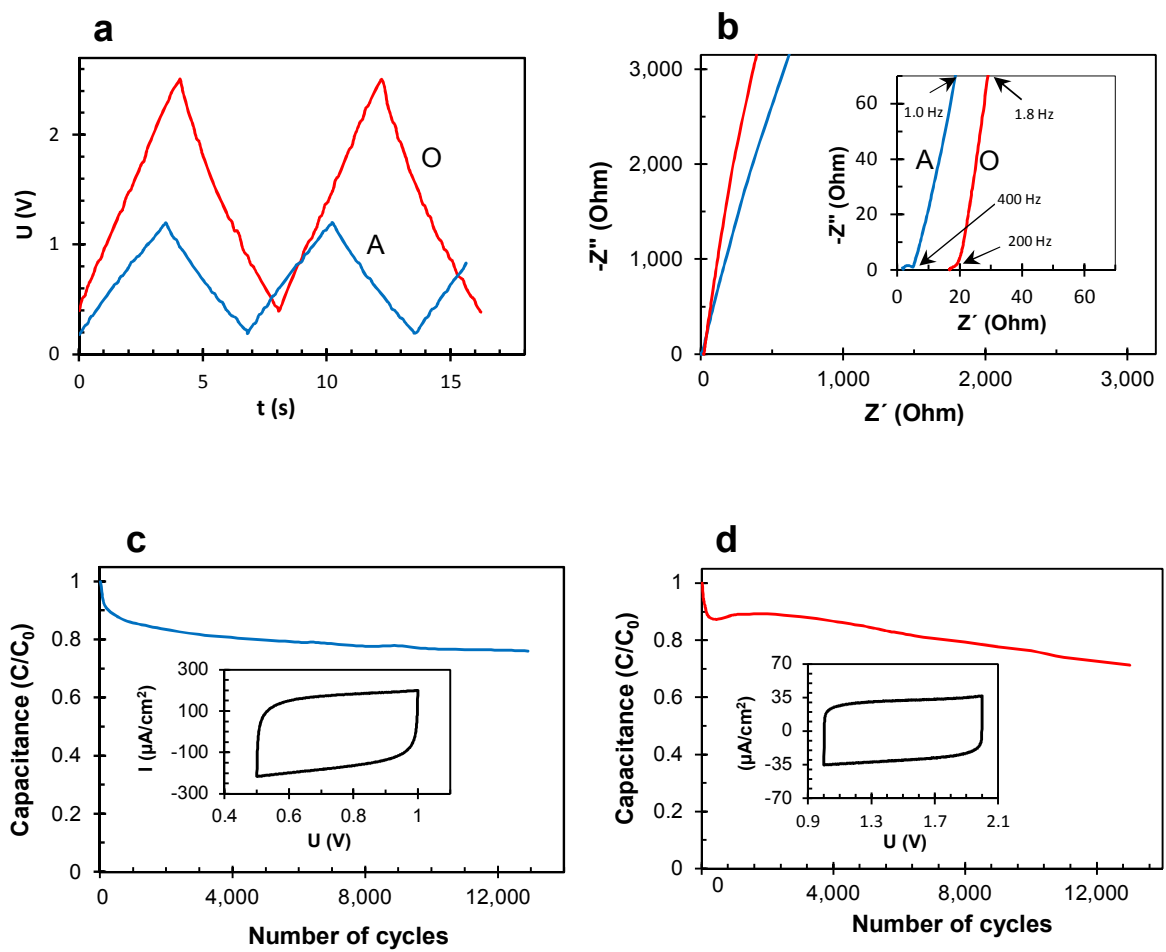

**Figure 4.** Electrochemical characteristics of the test assemblies of Figure 1 with aqueous (# A) and organic (# O) electrolyte. (a) Galvanostatic charge/discharge curves at 1.0 mA (corresponds to current density of 0.56 A/cm$^3$). (b) Nyquist plot obtained from electrochemical impedance spectroscopy measurements. Inset shows a high frequency part, with knee frequencies indicated (400 Hz and 200 Hz for samples with aqueous and organic electrolyte, respectively). (c, d) Capacitance retention during 13 000 cycles for device with aqueous (c) and organic (d) electrolyte. Insets show shape of cyclic voltammetry curves during the cycling.

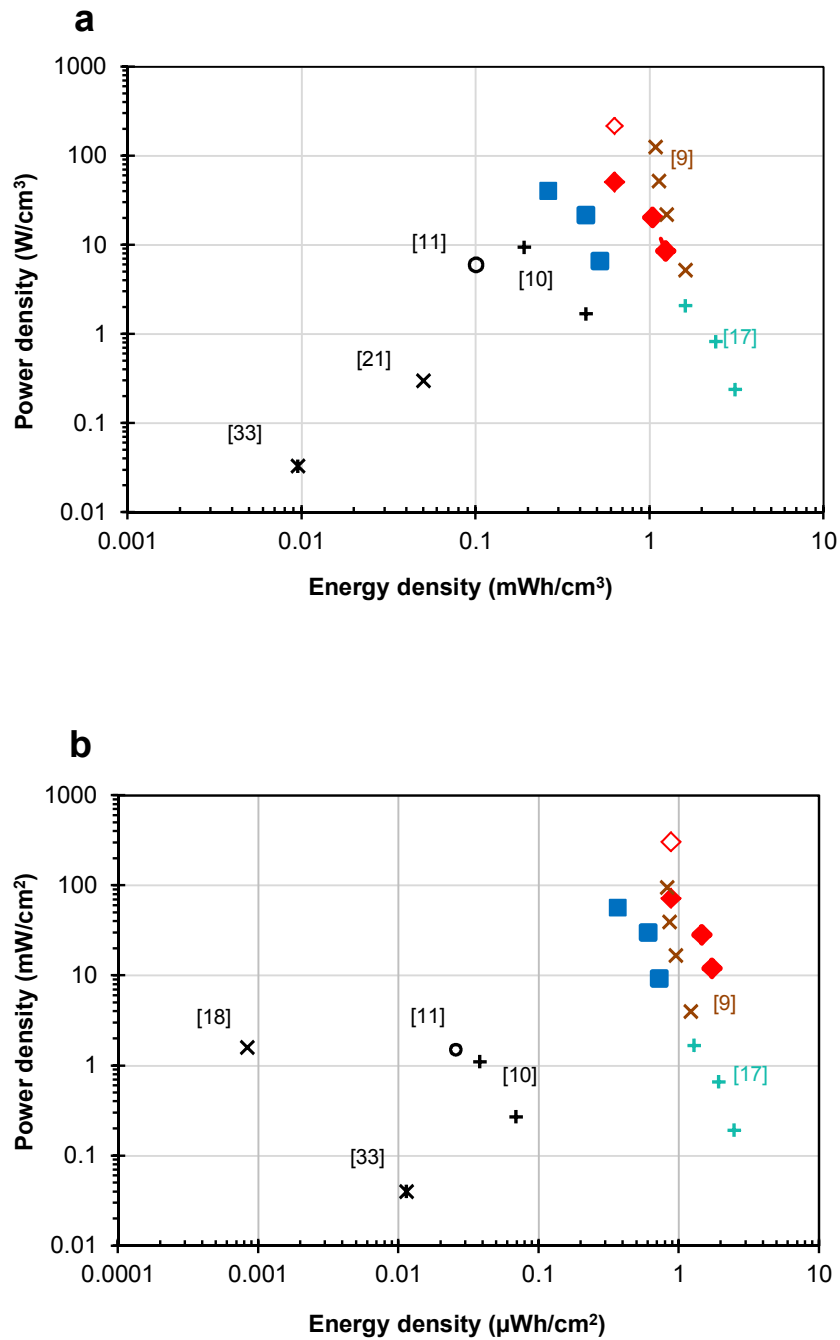

**Figure 5.** Ragone plot. (a) Volumetric and (b) areal energy and power densities of PS-TiN supercapacitors with aqueous (blue rectangulars) and organic (red diamonds) electrolyte and comparison with literature. Maximum values for 100 μm PDMS case with organic electrolyte are also shown (open red diamonds). The available data of several relevant devices from the literature are presented as well: laser graphene oxide [9], graphene coated porous silicon [17] and direct laser writing [10] (for those three cases the areal values were calculated from volumetric values using given device thicknesses); graphitization of silicon carbide [11] and silicon carbide nanowires [33] (volumetric values were calculated from areal values using device thickness); doped silicon nanowires [18] (device or layer thickness not given in the article); carbon fabrics with TiN nanowires [21].

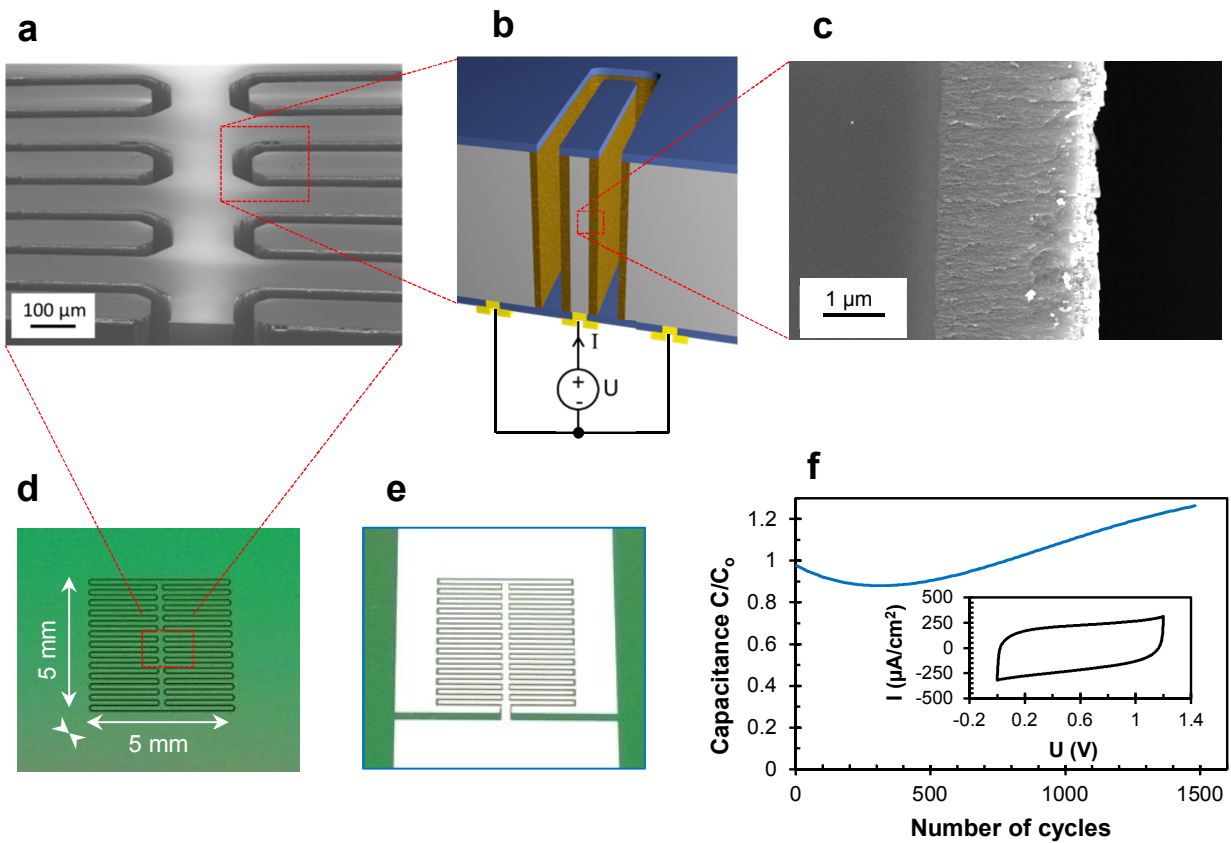

**Figure 6.** In-chip PS-TiN supercapacitor. (a) SEM picture (tilted view) of the trenches separating the electrodes. (b) Schematic illustration of the cross-section of two opposite electrodes of a ready device (TiN coated PS layer and the aluminium contact pads on the back side are also present). (c) Higher magnification SEM picture of the porous regions.(d) Device trench side and (e) the metallization side containing aluminum contacts for supercapacitor electrodes. (f) Cyclic voltammetry curve at 100 mV/s (inset) and capacitance retention.